\documentclass[12pt]{article}
\usepackage{amssymb,amsmath}
\usepackage{color}
\numberwithin{equation}{section}
\newtheorem{theorem}{Theorem}[section]

\newtheorem{proposition}[theorem]{Proposition}

\newtheorem{corollary}[theorem]{Corollary}

\newtheorem{remark}[theorem]{Remark}

\def\d{\partial}
\def\n{\noindent}
\def\f{\frac}

\def\proof{\noindent\hspace{2em}{\itshape Proof: }}
\def\QEDclosed{\mbox{\rule[0pt]{1.3ex}{1.3ex}}} 

\def\QED{\QEDclosed} 
\def\endproof{\hspace*{\fill}~\QED\par\endtrivlist\unskip}
\newcommand{\eqa}{\begin{eqnarray}}
\newcommand{\eeqa}{\end{eqnarray}}
\newcommand{\beq}{\begin{equation}}
\newcommand{\eeq}{\end{equation}}

\begin{document}
\title{Darboux-Egorov system,\\
bi-flat $F$-manifolds and Painlev\'e VI}
\author{Paolo Lorenzoni\\
\\
{\small Dipartimento di Matematica e Applicazioni}\\
{\small Universit\`a di Milano-Bicocca,}
{\small Via Roberto Cozzi 53, I-20125 Milano, Italy}\\
{\small paolo.lorenzoni@unimib.it}}

\date{}

\maketitle

\begin{abstract}
This is a generalization of the procedure presented in \cite{AL-DE} to construct semisimple bi-flat $F$-manifolds
 $(M,\nabla^{(1)},\nabla^{(2)},\circ,*,e,E)$ starting from homogeneous solutions of degree $-1$
 of Darboux-Egorov-system. The Lam\'e coefficients $H_i$
 involved in the construction are still homogeneous functions
  of a certain degree $d_i$ but we consider the general case $d_i\ne d_j$. As a consequence the rotation coefficients $\beta_{ij}$
 are homogeneous functions of degree $d_i-d_j-1$. It turns out that any semisimple bi-flat $F$ manifold satisfying a natural additional assumption  can be obtained in this way.
  Finally we show that  three dimensional semisimple bi-flat $F$-manifolds are parametrized by solutions of the full family of Painlev\'e VI. 
\end{abstract}

\section{Introduction}

A \emph{bi-flat} semisimple $F$-manifold $(M,\nabla^{(1)},\nabla^{(2)},\circ,*,e,E)$
 is a manifold $M$ endowed with a pair
 of flat connections $\nabla^{(1)}$ and $\nabla^{(2)}$, a pair of products $\circ$ and $*$ on the tangent spaces $T_u M$ and  a pair of vector fields $e$ and  $E$ satisfying the following conditions:
\begin{itemize}
\item the product $\circ$ is commutative, associative and with unity $e$.
 Moreover it is semisiple; this means that there exists a special set of coordinates, called canonical coordinates, such that the structure constants
 of $\circ$ reduce to the standard form $c^i_{jk}=\delta^i_j\delta^i_k$.
\item the product $*$ is also commutative, associative and with unity $E$. Moreover the operator $L=E\circ$ has vanishing Nijenhuis torsion and functionally independent eigenvalues. As a consequence, in canonical coordinates for $\circ$, the structure constants of $*$ read $c^{*i}_{jk}=\f{1}{E^i(u^i)}\delta^i_j\delta^i_k$.
\item $\nabla^{(1)}$ is compatible with the product $\circ$ and $\nabla^{(2)}$ is compatible with the product $*$:
\beq
\nabla^{(1)}_lc^i_{jk}=\nabla^{(1)}_jc^i_{lk},\qquad \nabla^{(2)}_lc^{*i}_{jk}=\nabla^{(2)}_jc^{*i}_{lk}
\eeq
\item $\nabla^{(1)} e=0$ and $\nabla^{(2)} E=0$,
\item $\nabla^{(1)}$ and $\nabla^{(2)}$ are almost hydrodynamically equivalent  i.e.
\beq\label{almostcomp}
(d_{\nabla^{(1)}}-d_{\nabla^{(2)}})(X\,\circ)=0,\qquad{\rm or}\qquad(d_{\nabla^{(1)}}-d_{\nabla^{(2)}})(X\,*)=0
\eeq
for every vector fields $X$; here $d_{\nabla}$ is the exterior covariant derivative 
 constructed from a connection $\nabla$.  
\end{itemize}

Bi-flat $F$-manifolds are a natural generalization of Frobenius manifolds. In the Frobenius case $\nabla^{(1)}$
 is the Levi-Civita connection of a metric $\eta$ which is invariant with respect to the product. This extra assumption 
 has two important consequences:

\n
- in flat coordinates for  $\nabla^{(1)}$, one has
$$\eta_{il}c^l_{jk}=\d_i\d_j\d_k F$$
for a suitable function $F$, called \emph{the Frobenius potential}.

\n
- the associated integrable hierarchy of PDEs, the \emph{principal hierarchy}, is Hamiltonian with respect 
 to the local Poisson bracket of hydrodynamic type defined by the metric $\eta$.

This means that, in general, the structure constants of  bi-flat $F$ manifolds do not admit any Frobenius
 potential and the associated integrable hierarchies are not Hamiltonian with respect to a local Poisson
 bracket of hydrodynamic type, at least  in the usual sense (they become Hamiltonian in a weaker sense
 if one considers local Poisson bracket on 1-forms \cite{AL-Poisson}).

In \cite{AL-DE} it was shown how to construct semisimple bi-flat $F$-manifolds starting from the solutions
 of the Darboux-Egorov system \cite{Darboux,Egorov}
\begin{eqnarray}
\label{ED1}
&&\d_k\beta_{ij}=\beta_{ik}\beta_{kj},\qquad k\ne i\ne j\ne k\\
\label{ED2}
&&e(\beta_{ij})=0,\\
\end{eqnarray}
augmented with the condition
\beq\label{ED3old}
E(\beta_{ij})=-\beta_{ij}.
\eeq
In the symmetric case $\beta_{ij}=\beta_{ji}$ the construction reduces to  the usual Dubrovin procedure
 to define semisimple Frobenius manifolds from solutions of Darboux-Egorov system. The non trivial point in
 the generalization is the relation between the connection $\nabla^{(1)}$ and the Lam\'e coefficients $H_i$ involved in the construction: in the non symmetric case 
  the connection $\nabla^{(1)}$ is no longer  the Levi-Civita connection of the diagonal
 metric $\eta_{ii}=H_i^2$. 

In the present paper we further extend Dubrovin procedure considering instead of
 \eqref{ED3old} the more general condition 
\beq
\label{ED3}
E(\beta_{ij})=(d_i-d_j-1)\beta_{ij}.
\eeq
This adds $n-1$ free parameters to the theory.  Remarkably, in the case $n=3$ the system 
 (\ref{ED1},\ref{ED2},\ref{ED3}) is equivalent to the full family 
 of Painlev\'e VI (a more precise statement will be given in Section 5).  
Notice that the additional constraint \eqref{ED3} is not compatible with $\beta_{ij}=\beta_{ji}$ since 
$$E(\beta_{ij})-E(\beta_{ji})=2(d_i-d_j)\beta_{ij}$$
and therefore the case $d_i\ne d_{j}$ does not produce new examples of Frobenius manifolds.

The paper is organized as follows. In Section 2 we show how to construct
 bi-flat $F$ manifolds starting from solutions of (\ref{ED1},\ref{ED2},\ref{ED3}). We also show that if we assume
 that the eigenvalues of $E\circ$ are canonical coordinates, then all bi-flat $F$ manifolds can be obtained in this way.
 The case $n=2$ and $n=3$ are treated in Section 3 and 4. Section 4 is also devoted to discuss how the solutions of the system
 (\ref{ED1},\ref{ED2},\ref{ED3}) are related to the sigma form of Painlev\'e VI. In the final Section 5 we discuss an example.

\section{From Darboux-Egorov system to
 bi-flat $F$ manifolds}
From now on we will work in canonical coordinates $(u^1,\dots,u^n)$ and we will denote by $\d_i$ the partial derivative $\f{\d}{\d u^i}$.
 Moreover by definition $e=\sum_{i=1}^n\d_i$ and $E=\sum_{i=1}^n u^i\d_i$. 
\begin{theorem}\label{mainth}
Let $\beta_{ij}$ be a solution of the system  (\ref{ED1},\ref{ED2},\ref{ED3}) and $(H_1,\dots,H_n)$
 a solution of the system 
\begin{eqnarray}
\label{L1}
&&\d_j H_i=\beta_{ij}H_j,\qquad i\ne j\\
\label{L2}
&&e(H_i)=0,
\end{eqnarray}
 satisfying
 the condition 
\begin{equation}
\label{L3}
E(H_i)=d_iH_i,
\end{equation} 
then  
\begin{itemize}
\item the natural connection $\nabla_1$ defined by
\begin{equation}\label{naturalc}
\begin{split}
\Gamma^i_{jk}&:=0\qquad\forall i\ne j\ne k \ne i\\
\Gamma^i_{jj}&:=-\Gamma^i_{ij}\qquad i\ne j\\
\Gamma^i_{ij}&:=\f{H_j}{H_i}\beta_{ij}\qquad i\ne j\\
\Gamma^i_{ii}&:=-\sum_{l\ne i}\Gamma^i_{li},
\end{split}
\end{equation} 
\item the dual connection $\nabla_2$ defined by
\begin{equation}\label{dualnabla}
\begin{split}
\Gamma^i_{jk}&:=0\qquad\forall i\ne j\ne k \ne i\\
\Gamma^i_{jj}&:=-\f{u^i}{u^j}\Gamma^i_{ij}\qquad i\ne j\\
\Gamma^i_{ij}&:=\f{H_j}{H_i}\beta_{ij}\qquad i\ne j\\
\Gamma^i_{ii}&:=-\sum_{l\ne i}\f{u^l}{u^i}\Gamma^i_{li}-\f{1}{u^i},
\end{split}
\end{equation}
\item the structure constants  defined in the coordinates $(u^1,\dots,u^n)$ by $c^i_{jk}=\delta^i_j\delta^i_k$,
\item the structure constants  defined in the coordinates $(u^1,\dots,u^n)$ by $c^{*i}_{jk}=\f{1}{u^i}\delta^i_j\delta^i_k$, 
\item the vector fields $e$ and $E$,  
\end{itemize}
define a bi-flat semisimple $F$-manifold $(M,\nabla_1,\nabla_2,\circ,*,e,E)$. 
\end{theorem}

\n
\emph{Proof}. The flatness of the connections $\nabla^{(1)}$ and $\nabla^{(2)}$
 can be proved by straightforward computation. Moreover,
by construction, the connection $\nabla_1$ defined in \eqref{naturalc} is compatible with the product $c^i_{jk}=\delta^i_j\delta^i_k$ and satisfies $\nabla_1 e=0$ and the connection $\nabla_2$  defined
 in \eqref{dualnabla} is compatible with the product $c^{*i}_{jk}=\f{\delta^i_j\delta^i_k}{u^i}$
 and satisfies $\nabla_2 E=0$. 

Finally, the natural connection and the dual connection
 associated to the same functions $H_i$ are almost hydrodynamically equivalent by definition since 
$$\Gamma^{(1)i}_{ij}=\Gamma^{(2)i}_{ij}=\f{H_j}{H_i}\beta_{ij}.$$

\endproof

A natural question arises: 
does any bi-flat $F$-manifold come from a solution of the system (\ref{ED1},\ref{ED2},\ref{ED3},\ref{L1},\ref{L2},\ref{L3})?
 The answer is given by the following theorem.
 \begin{theorem}
Let  $(M,\nabla^{(1)},\nabla^{(2)},\circ,*,e,E)$ be a bi-flat $F$-manifold such that the  
 eigenvalues of $E\circ$ are canonical coordinates. Then there exist $(H_i,\beta_{ij})$ satisfying
 the system (\ref{ED1},\ref{ED2},\ref{ED3},\ref{L1},\ref{L2},\ref{L3}) such that, in canonical coordinates
$$\Gamma^{(1)i}_{ij}=\Gamma^{(2)i}_{ij}=\f{H_j}{H_i}\beta_{ij}.$$ 
\end{theorem}

\proof
In canonical coordinates $\nabla^{(1)}$ is given by  \eqref{naturalc} and $e=\sum_l\f{\d}{\d u^l}$. Moreover, 
due to the additional assumption in canonical coordinates $E=\sum_l u^l\f{\d}{\d u^l}$ 
  and $\nabla^{(2)}$ is given by \eqref{dualnabla}. Since $\nabla^{(1)}$ and $\nabla^{(2)}$ are almost hydrodynamically equivalent 
 we have also
$$\Gamma^{(1)i}_{ij}=\Gamma^{(2)i}_{ij}:=\Gamma^i_{ij},\qquad\forall i\ne j.$$ 
Now we have to exploit the flatness of $\nabla^{(1)}$ and $\nabla^{(2)}$. From
$$R^{(1)i}_{ikj}=R^{(2)i}_{ikj}=\d_k\Gamma^{i}_{ij}-\d_j\Gamma^{i}_{ik}=0,$$
it folllows that there exist $H_i$ such that
$$\Gamma^{i}_{ij}=\d_j\ln{H_i}$$
Clearly $H_i$ is defined up to a multiplicative factor depending only on $u^i$. 
Using $R^{(1)i}_{iji}=0$ and $R^{(1)i}_{ijl}=0$ we obtain
$$e(\Gamma^i_{ij})=\d_i\Gamma^{i}_{ij}+\sum_{l\ne i}\d_l\Gamma^{i}_{ij}=
\d_j\Gamma^{(1)i}_{ii}+\sum_{l\ne i}\d_l\Gamma^{i}_{ij}=
-\sum_{l\ne i}\d_j\Gamma^{i}_{il}+\sum_{l\ne i}\d_j\Gamma^{i}_{il}=0.$$
This implies $\d_j\left(\f{e(H_i)}{H_i}\right)=0$, that is $e(H_i)=c_i(u^i)H_i$. Due to the freedom in the choice 
 of $H_i$, without loss of generality we can assume $c^i=0$.  Similarly, using the flatness of the dual connection (in particular
 $R^{(2)i}_{iji}=0$ and $R^{(2)i}_{ijl}=0$) we obtain
\begin{eqnarray*}
E(\Gamma^i_{ij})&=&u^i\d_i\Gamma^i_{ij}+\sum_{l\ne i}u^l\d_l\Gamma^i_{ij}=
u^i\d_j\Gamma^{(2)i}_{ii}+\sum_{l\ne i}u^l\d_l\Gamma^i_{ij}=\\
&&-\sum_{l\ne i}\d_j\left(u^l\Gamma^i_{il}\right)+\sum_{l\ne i}u^l\d_j\Gamma^i_{il}=
-\Gamma^i_{ij}
\end{eqnarray*}
and, as a consequence:
$$\d_j\left(E(\ln{H_i})\right)=E\left(\d_j\ln{H_i}\right)+\d_j\ln{H_i}=0,\qquad\forall j\ne i.$$
This means that $E(H_i)=d_i(u^i)H_i$. We have to prove $\d_i d_i=0$. By straightforward
 computation we obtain
\begin{eqnarray*}
\d_i d_i&=&\d_i\left(\f{E(H_i)}{H_i}\right)=\f{E(\d_i H_i)+\d_i H_i}{H_i}-\f{E(H_i)\d_i H_i}{H_i^2}=\\
&&\f{E\left(-\sum_{l\ne i}\d_l H_i\right)-\sum_{l\ne i}\d_l H_i+d^i\sum_{l\ne i}\d_l H_i}{H_i}=\\
&&\f{-\sum_{l\ne i}\d_l(E(H_i))+d_i\sum_{l\ne i}\d_l H_i}{H_i}=0.
\end{eqnarray*}
Let us define the rotation coefficients as 
$$\beta_{ij}=\f{\d_j H_i}{H_j}=\f{H_i}{H_j}\Gamma^i_{ij}.$$ 
It remains to prove \eqref{ED2}, \eqref{ED3} and \eqref{ED1}. Due to $e(\beta_{ij})=0$,
 $E(H_i)=d_iH_i$ and $E(\Gamma^i_{ij})=-\Gamma^i_{ij}$, the first and the second ones are  elementary. 
  The last one follows from $R^{(1)i}_{jki}=R^{(2)i}_{jki}=0$:
\begin{eqnarray*}
&&0=\d_k\Gamma^i_{ij}+\Gamma^i_{ik}\Gamma^i_{ij}-\Gamma^i_{ij}\Gamma^j_{jk}
-\Gamma^i_{ik}\Gamma^k_{kj}=\\
&&\d_k\left(\f{H_j}{H_i}\beta_{ij}\right)+\f{H_k H_j}{(H_i)^2}\beta_{ik}\beta_{ij}-\f{H_k}{H_i}\beta_{ij}\beta_{jk}
-\f{H_j}{H_i}\beta_{ik}\beta_{kj}=\\
&&\f{\d_k H_j}{H_i}\beta_{ij}-\f{H_j\d_k H_i}{(H_i)^2}\beta_{ij}+\f{H_j}{H_i}\d_k\beta_{ij}+\f{H_k H_j}{(H_i)^2}\beta_{ik}\beta_{ij}-\f{H_k}{H_i}\beta_{ij}\beta_{jk}
-\f{H_j}{H_i}\beta_{ik}\beta_{kj}=\\
&&\f{H_j}{H_i}\left(\d_k\beta_{ij}-\beta_{ik}\beta_{kj}\right).
\end{eqnarray*}

\endproof

\begin{remark}
Both the systems (\ref{ED1},\ref{ED2},\ref{ED3}) and  (\ref{L1},\ref{L2}) (given $\beta_{ij}$ satisfying  (\ref{ED1},\ref{ED2}))
 are compatible. The proof is a straightforward (not short) computation. For arbitary values of the constant $d$, system (\ref{L1},\ref{L2},\ref{L3}) 
  does not admit solutions. The choice of the right degrees of homogeneity can be done adapting the  procedure used by Dubrovin in \cite{du93} for the symmetric case. 
\newline
\newline
The key observation is that the system (\ref{ED1},\ref{ED2},\ref{ED3}) can be written in the  Lax form
\footnote{by definition $\beta_{ii}=0$.}
$$\d_k V=[V,W]$$
where $V_{ij}=(u^j-u^i)\beta_{ij}-(d_j-d_1)\delta^i_j$ and $W_{ij}=\delta^k_i\beta_{kj}-\beta_{ik}\delta^k_j$ (clearly instead of $d_1$ we can choose $d_2,\dots,d_n$). 
Moreover the system 
  (\ref{L1},\ref{L2})  is equivalent to
$$\d_k H=-WH$$
where $H=(H_1,\dots,H_n)$. 
Using these facts  it is easy to check that 
\begin{itemize}
\item the matrix $V$ acts on the space of solutions of the linear system (\ref{L1},\ref{L2}), 
\item the eigenvalues
 of $V$ do not depend on $u$. 
\item $d_1$ must be an eigenvalue of $V$. Indeed
 the eigenvectors $H^{(\alpha)}=(H_1^{(\alpha)},\dots,H_n^{(\alpha)})$ of $V$ satisfy the equation:
$$E(H^{(\alpha)}_i)=(d_i-d_1+\mu)H^{(\alpha)}_i.$$
\end{itemize}
\end{remark}

\section{Examples in the case $n=2$}
In this case the Egorov-Darboux system reduces to
\begin{eqnarray*}
&&\f{\d\beta_{ij}}{\d u^1}+\f{\d\beta_{ij}}{\d u^2}=0,\\
&&u^1\f{\d\beta_{ij}}{\d u^1}+u^2\f{\d\beta_{ij}}{\d u^2}=(d_i-d_j-1)\beta_{ij}.
\end{eqnarray*}
The first equations tell us that the rotation coefficients depend only on the difference $(u^1-u^2)$. The remaining equations tell us that they
 are homogeneous
 functions of degree $-1$. This gives us
\begin{eqnarray*}
\beta_{12}&=&C_1(u^1-u^2)^{d_1-d_2-1},\\
\beta_{21}&=&C_2(u^1-u^2)^{d_2-d_1-1}.
\end{eqnarray*}
To construct the natural connections we need to solve the system
 for the Lam\'e coefficients:
\begin{eqnarray*}
&&\f{\d H_i}{\d u^1}+\f{\d H_i}{\d u^2}=0,\\
&&u^1\f{\d H_i}{\d u^1}+u^2\f{\d H_i}{\d u^2}=d_iH_i,\\
&&\d_2 H_1=C_1(u^1-u^2)^{d_1-d_2-1}H_2,\\
&&\d_1 H_2=C_2(u^1-u^2)^{d_2-d_1-1}H_1.
\end{eqnarray*}
The first two equations imply
\begin{eqnarray*}
H_1&=&D_1(u^1-u^2)^{d_1},\\
H_2&=&D_2(u^1-u^2)^{d_2}.
\end{eqnarray*}
Due to the remaining equations the constants $D_1, D_2,d_1,d_2$ obbey two additional additional constraints:
$$-d_1D_1=C_1D_2$$
and
$$d_2D_2=C_2D_1.$$
Multiplying both equations we obtain
\beq\label{constraint}
d_1d_2=-C_1C_2.
\eeq
The same result can be obtained computing the eigenvalues of the matrix $V$ 
\beq
\begin{pmatrix}
0 & -C_1\\
C_2 & d_1-d_2
\end{pmatrix}.
\eeq
We have
$$\lambda=\frac{d_1-d_2
\pm\sqrt{(d_1-d_2)^2-4C_1C_2}}{2}.$$
If we impose that $d_1$ is an eigenvalue we obtain
 the constraint \eqref{constraint}.

For any choice of $C_1$ and $C_2$ 
the natural and dual connections $\nabla_1$ and $\nabla_2$ are defined by \eqref{naturalc} and \eqref{dualnabla} with
\begin{eqnarray*}
\Gamma^1_{12}&=&\Gamma^{(1)1}_{12}=\Gamma^{(1)}_{12}=\f{D_2}{D_1}\f{C_1}{u^1-u^2}=\f{d_1}{u^2-u^1},\\
\Gamma^2_{21}&=&\Gamma^{(2)1}_{12}=\Gamma^{(2)}_{12}=\f{D_1}{D_2}\f{C_2}{u^1-u^2}=\f{d_2}{u^1-u^2}.
\end{eqnarray*}

\section{Bi-flat $F$-manifolds in dimension $n=3$}

In this Section 
we show that the system (\ref{ED1},\ref{ED2},\ref{ED3}) is equivalent to the sigma form of Painlev\'e VI.
  In literature, the relation between Darboux-Egorov (or the related $N$-wave system) and Painlev\'e VI has been  studied by several authors
 (for instance \cite{FLMS,K,KK,CGM,AVdL}).The proof we present
 here  is elementary. In one direction (from Darboux-Egorov to  Painlev\'e VI) it is based on \cite{AVdL}.
 In the other direction we extend the proof given in
 \cite{AL-DE} in the case $d_i=d_j$.
\newline
\newline
First of all, we observe that, 
due to \eqref{ED2} and  \eqref{ED3}, the rotation coefficients $\beta_{ij}$ are homogeneous functions of degree
 $d_i-d_j-1$ depending only on the difference of the coordinates. Without loss of generality we can 
 write them in the form
\begin{equation}\label{relationsbetaF}
\begin{split}
\beta_{12}=\frac{1}{u^2-u^1}F_{12}\left(\frac{u^3-u^1}{u^2-u^1} \right)(u^2-u^1)^{d_1-d_2}\\
\beta_{21}=\frac{1}{u^2-u^1}F_{21}\left(\frac{u^3-u^1}{u^2-u^1} \right)(u^2-u^1)^{d_2-d_1}\\
\beta_{32}=\frac{1}{u^3-u^2}F_{32}\left(\frac{u^3-u^1}{u^2-u^1} \right)(u^2-u^1)^{d_3-d_2}\\
\beta_{23}=\frac{1}{u^3-u^2}F_{23}\left(\frac{u^3-u^1}{u^2-u^1} \right)(u^2-u^1)^{d_2-d_3}\\
\beta_{13}=\frac{1}{u^3-u^1}F_{13}\left(\frac{u^3-u^1}{u^2-u^1} \right)(u^2-u^1)^{d_1-d_3}\\
\beta_{31}=\frac{1}{u^3-u^1}F_{31}\left(\frac{u^3-u^1}{u^2-u^1} \right)(u^2-u^1)^{d_3-d_1}\\
\end{split}
\end{equation}
Putting $\eqref{relationsbetaF}$ into the system \eqref{ED1} 
 we obtain the system \eqref{finalODE3} for the functions $F_{ij}$.
\begin{equation}\label{finalODE3}
\begin{split}
  \frac{d}{dz} F_{12}&=\frac{1}{z(z-1)}F_{13}F_{32}\\
 \frac{d}{dz} F_{13}&=-\frac{1}{z-1}F_{12}F_{23}+\f{d_1-d_3}{z}F_{13}\\
\frac{d}{dz} F_{21}&=\frac{1}{z(z-1)}F_{23}F_{31}\\
 \frac{d}{dz} F_{23}&=\frac{1}{z}F_{21}F_{13}+\f{d_2-d_3}{z-1}F_{23}\\
\frac{d}{dz} F_{31}&=-\frac{1}{z-1}F_{32}F_{21}+\f{d_3-d_1}{z}F_{31}\\
\frac{d}{dz} F_{32}&=\frac{1}{z}F_{31}F_{12}+\f{d_3-d_2}{z-1}F_{32},\\
\end{split}
\end{equation}
where the independent variable $z:=\frac{u^3-u^1}{u^2-u^1}$.  

Now we discuss how the non-autonomous systems of ODEs \eqref{finalODE3} for the $F_{ij}$ can be reduced to the sigma form of Painlev\'e VI.
\begin{theorem}\label{painleveVI}
System \eqref{finalODE3} is equivalent to the following   equation:
\begin{equation}\label{painlevevi2}
\begin{split} 
&z^2(z-1)^2(f'')^2+4\left[f'(zf'-f)^2
-(f')^2(zf'-f) \right]-
(2R^2+d_{13}^2)(f')^2-d_{21}^2\left(zf'-f\right)^2+\\
&-2d_{21}d_{13}f'\left(zf'-f\right)-((d_{13}+d_{23})R^2+2D)d_{21}\left(zf'-f\right)+\\
&-[((d_{13}+d_{23})R^2+2D)d_{13}+R^4]f'
-\left(D+\f{(d_{13}+d_{23})R^2}{2}\right)^2=0
\end{split}
\end{equation}
After the substitution $f=\psi+az=\phi=az+b$ with $a=\f{d_{21}^2}{4}$ and $b=-\f{d_{21}d_{23}}{4}$
the equation \eqref{painlevevi2} reduces to
\begin{equation}\label{painlevevi3}
\begin{split}
&z^2(z-1)^2(\phi'')^2+4\left[\phi'(z\phi'-\phi)^2
-(\phi')^2(z\phi'-\phi) \right]-
\left[2R^2+d_{13}^2+d_{21}d_{23}\right](\phi')^2+\\
&-\left[2Dd_{21}
+(d_{13}d_{21}+d_{23}d_{21})R^2+\f{d_{21}^4}{4}+\f{d_{21}^3d_{13}}{2}\right](z\phi'-\phi)+\\
&-\left[R^4+2Dd_{13}+((d_{21}^2+d^2_{13}+d_{23}d_{13})R^2-
\f{d_{21}^2d_{23}^2}{4}+\f{d_{21}^2d_{13}d_{23}}{2}+\f{d_{21}^3d_{23}}{2}+\f{d_{21}^2d_{13}^2}{2}\right]\phi'+\\
&-D^2-DR^2(d_{13}+d_{23})-\f{R^4}{4}\left[d_{21}^2+(d_{13}+d_{23})^2\right]
-\f{D}{2}d_{21}^2(d_{13}+d_{23})+\\
&-\f{R^2}{8}d_{21}^2\left[d_{21}^2
+4d_{13}d_{23}+2d_{13}^2
+2d_{23}^2\right]-\f{d_{21}^4}{16}\left[d_{23}d_{21}
+d_{13}^2+2d_{13}d_{23}\right]
\end{split}
\end{equation}
which is the sigma form of Painlev\'e VI equation:
\begin{equation}\label{sigmapainleve2}
\begin{split}
z^2(z-1)^2(\sigma'')^2+4\left[\sigma'(z\sigma'-\sigma)^2-(\sigma')^2(z\sigma'-\sigma) \right]-4v_1v_2v_3v_4(z\sigma'-\sigma)+\\
-(\sigma')^2\left(\sum_{k=1}^4 v_k^2 \right)-\sigma' \left(\sum_{i<j}^4v_{i}^2v_{j}^2-2v_1v_2v_3v_4 \right)-\sum_{i<j<k}^4v_i^2v_j^2v_k^2.
\end{split}
\end{equation}
where the parameters $v^2_1, v^2_2, v^2_3,v^2_4$ are the roots of the polynomial 
\begin{equation}\label{polynomial}
\begin{split}
&\lambda^4-(2R^2+d_{13}^2-d_{21}d_{13})\lambda^3+\\
&+\left[R^4+D(2d_{13}+d_{21})+
\left(\f{d_{13}d_{21}}{2}+\f{d_{23}d_{21}}{2}+d_{21}^2+d^2_{13}+d_{23}d_{13}\right)R^2+\right.\\
&\left.-\f{d_{21}^2d_{23}^2}{4}+\f{d_{21}^4}{8}+\f{d_{21}^3d_{13}}{4}+\f{d_{21}^2d_{13}d_{23}}{2}+\f{d_{21}^3d_{23}}{2}+\f{d_{21}^2d_{13}^2}{2}\right]\lambda^2+\\
&-\left[D^2+DR^2(d_{13}+d_{23})-\f{R^4}{4}\left(d_{21}^2+(d_{13}+d_{23})^2\right)
+\f{D}{2}d_{21}^2(d_{13}+d_{23})+\right.\\
&\left.+\f{R^2}{8}d_{21}^2\left(d_{21}^2
+4d_{13}d_{23}+2d_{13}^2
+2d_{23}^2\right)+\f{d_{21}^4}{16}\left(d_{23}d_{21}
+d_{13}^2+2d_{13}d_{23}\right)\right]\lambda+\\
&+\left[\f{D}{2}d_{21}
+\f{R^2}{4}(d_{13}d_{21}+d_{23}d_{21})+\f{d_{21}^4}{16}+\f{d_{21}^3d_{13}}{8}\right]^2.
\end{split}
\end{equation}
\end{theorem}

\n
\emph{Proof}. By straightforward computation we get 
\begin{eqnarray*}
&&\f{d}{dz}(F_{12}F_{21}+F_{13}F_{31}+F_{23}F_{32})=0
\end{eqnarray*}
and
\begin{eqnarray*}
&&\f{d}{dz}(F_{23}F_{31}F_{12}-F_{13}F_{32}F_{21}+d_{23}F_{13}F_{31}
+d_{13}F_{23}F_{32})=0
\end{eqnarray*}
where $d_{ij}:=d_i-d_j$. This implies
\beq\label{I1}
F_{12}F_{21}+F_{13}F_{31}+F_{23}F_{32}=-R^2
\eeq
and
\beq\label{I2}
F_{23}F_{31}F_{12}-F_{13}F_{32}F_{21}+d_{23}F_{13}F_{31}
+d_{13}F_{23}F_{32}=D
\eeq
for some constants $R$ and $D$.

Let us introduce a function $f$ defined, up to a constant, by 
 \begin{equation}\label{position12}F_{12}F_{21}:=f'
\end{equation}
Due to equations \eqref{finalODE3}, we have 
\begin{eqnarray*}
&&\frac{d}{dz}\left(F_{13}F_{31} \right)=F'_{13}F_{31}+F_{13}F'_{31}=\\
&&=-\frac{1}{z-1}F_{12}F_{23}F_{31}+\f{d_1-d_3}{z}F_{13}F_{31}
-\frac{1}{z-1}F_{32}F_{21}F_{13}+\f{d_3-d_1}{z}F_{13}F_{31}=\\
&&=-z\frac{d}{dz}\left(F_{12}F_{21}\right)=F_{12}F_{21}-\frac{d}{dz}\left(zF_{12}F_{21} \right)=\f{d}{dz}\left(f-zf'\right)
\end{eqnarray*}
Thus, choosing the integration constant equal to $-\f{R^2}{2}$ we have
\begin{equation}\label{position13}
F_{13}F_{31}=f-zf'-\f{R^2}{2}.
\end{equation}
and consequently
\begin{equation}\label{position23}
F_{23}F_{32}=-R^2-F_{12}F_{21}-F_{13}F_{31}=(z-1)f'-f-\f{R^2}{2}.
\end{equation}
We want to derive a second order ODE for the function $f$. This can be easily done writing the second derivative of $f$ in terms of  the products $F_{12}F_{21}$, 
 $F_{13}F_{31}$ and $F_{23}F_{32}$. We have
\begin{eqnarray*}
&&[z(z-1)f'']^2=\left[z(z-1)\frac{d}{dz}\left(F_{12}F_{21}\right)\right]^2
=[F_{21}F_{13}F_{32}+F_{12}F_{31}F_{23}]^2\\
&&4\left( F_{12}F_{21}F_{13}F_{31}F_{23}F_{32}\right)+(D-d_{23}F_{13}F_{31}
-d_{13}F_{23}F_{32})^2=\\
&&4f'g_1g_2+\left[D-d_{23}g_1
-d_{13}g_2\right]^2,
\end{eqnarray*}
where  $g_1=f-zf'-\f{R^2}{2}$,  $g_2=-f+(z-1)f'-\f{R^2}{2}$. Expanding the above expression, 
after some computations one obtains the equation
 \eqref{painlevevi2}.
\newline
This proves that given a solution of system \eqref{finalODE3} we can construct a solution of \eqref{painlevevi2}.
\newline
\newline
Viceversa given any solution $f$ of \eqref{painlevevi2}
 the corresponding solution $F_{ij}$ of \eqref{finalODE3} is defined by
\begin{equation}\label{Fij}
\begin{split}
F_{12}&=\sqrt{f'}\;\exp\left({- \int_{z_0}^z\left[\frac{\varphi}{2t(t-1)f'}\right]\,dt+C_{12}}\right),\\
F_{21}&=\sqrt{f'}\;\exp\left({\int_{z_0}^z\left[\frac{\varphi}{2t(t-1)f'}\right]\,dt+C_{21}}\right),\\
F_{13}&=\sqrt{g_1}\;\exp\left({-\int_{z_0}^z\left[\frac{\varphi}{2(t-1)g_1}
-\f{d_{13}}{t}\right]\,dt+C_{13}}\right),\\
F_{31}&=\sqrt{g_1}\;\exp\left({\int_{z_0}^z\left[\frac{\varphi}{2(t-1)g_1}
-\f{d_{13}}{t}\right]\,dt+C_{31}}\right),\\
F_{23}&=\sqrt{g_2}\;\exp\left({-\int_{z_0}^z\left[\frac{\varphi}{2t\,g_2}-\f{d_{23}}{t-1}\right]\,dt+C_{23}}\right),\\
F_{32}&=\sqrt{g_2}\;\exp\left({\int_{z_0}^z\left[\frac{\varphi}{2t\,g_2}-\f{d_{23}}{t-1}\right]\,dt+C_{32}}\right), 
\end{split}
\end{equation}
where $\varphi=D-d_{23}g_1-d_{13}g_2$ and  $C_{ij}$ are integration constants
 satisfying the linear system
 \begin{equation}\label{Cij}
\begin{split}
&-C_{12}+C_{13}+C_{32}-\ln{(f''(z_0)z_0(z_0-1)-\varphi(z_0))}+\ln{(2\sqrt{f'(z_0)g_1(z_0)
g_2(z_0)})}=0\cr
&-C_{21}+C_{23}+C_{31}-\ln{(f''(z_0)z_0(z_0-1)+\varphi(z_0))}+\ln{(2\sqrt{f'(z_0)g_1(z_0)
g_2(z_0)})}=0\cr
&-C_{13}+C_{12}+C_{23}-\ln{(f''(z_0)z_0(z_0-1)+\varphi(z_0))}+\ln{(2\sqrt{f'(z_0)g_1(z_0)
g_2(z_0)})}=0\cr
&-C_{31}+C_{32}+C_{21}-\ln{(f''(z_0)z_0(z_0-1)-\varphi(z_0))}+\ln{(2\sqrt{f'(z_0)g_1(z_0)
g_2(z_0)})}=0\cr
&-C_{23}+C_{21}+C_{13}-\ln{(f''(z_0)z_0(z_0-1)-\varphi(z_0))}+\ln{(2\sqrt{f'(z_0)g_1(z_0)
g_2(z_0)})}=0\cr
&-C_{32}+C_{31}+C_{12}-\ln{(f''(z_0)z_0(z_0-1)+\varphi(z_0))}+\ln{(2\sqrt{f'(z_0)g_1(z_0)
g_2(z_0)})}=0\\  
\end{split}
\end{equation}
The proof is a generalization of the proof given 
 in \cite{AL-DE} in the case $d_{ij}=0$ ($\varphi=D$). Substituting \eqref{Fij} in \eqref{finalODE3}, after some 
 computations we obtain
\begin{eqnarray*}
&&-C_{12}+C_{13}+C_{32}-\ln{(f''(z_0)z_0(z_0-1)-\varphi(z_0))}+\ln{(2\sqrt{f'(z_0)g_1(z_0)
g_2(z_0)})}+\\
&&-\int_{z_0}^z\f{d}{dt}\ln{(t(t-1)f''-\varphi)}\,dt+\int_{z_0}^z\f{d}{dt}\ln{[2\sqrt{f'g_1g_2}]}\,dt+\\
&&+\int_{z_0}^z\varphi\f{g_1g_2+(t-1)f'g_1-tf'g_2}{2t(t-1)f'g_1g_2}\,dt-\int_{z_0}^z\left[\f{d_{23}}{t-1}-\f{d_{13}}{t}\right]\,dt
=0,
\end{eqnarray*}
or, equivalently,
\begin{eqnarray*}	
&&-C_{12}+C_{13}+C_{32}-\ln{(f''(z_0)z_0(z_0-1)-\varphi(z_0))}+\ln{(2\sqrt{f'(z_0)g_1(z_0)
g_2(z_0)})}+\\
&&-\int_{z_0}^z\f{t(t-1)f'''+(2t-1)f''-\varphi'}{t(t-1)f''-\varphi}\,dt+\int_{z_0}^z\f{(t(t-1)f''+\varphi)\f{d}{dt}[f'g_1g_2]}{2t(t-1)f'g_1g_2f''}\, dt+\\
&&-\int_{z_0}^z\left[\f{d_{23}t-d_{13}(t-1)}{t(t-1)}\right]\,dt=0
\end{eqnarray*}
Using the equation \eqref{painlevevi2} written in the form
$$f'g_1g_2=\f{1}{4}(z(z-1)f''+\varphi)(z(z-1)f''-\varphi)$$
and the equation obtained from \eqref{painlevevi2} by differentiating with respect to $z$, we obtain
\begin{eqnarray*}
&&-C_{12}+C_{13}+C_{32}-\ln{(f''(z_0)z_0(z_0-1)-\varphi(z_0))}+\ln{(2\sqrt{f'(z_0)g_1(z_0)
g_2(z_0)})}+\\
&&-\int_{z_0}^z\f{t(t-1)f'''+(2t-1)f''-\varphi'}{t(t-1)f''-\varphi}\,dt
+\int_{z_0}^z\f{2\f{d}{dt}[f'g_1g_2]}{t(t-1)(t(t-1)f''-\varphi)f''}\,dt+\\
&&-\int_{z_0}^z\left[\f{\varphi'}{t(t-1)f''}\right]\,dt=\\
&&-C_{12}+C_{13}+C_{32}-\ln{(f''(z_0)z_0(z_0-1)-\varphi(z_0))}+\ln{(2\sqrt{f'(z_0)g_1(z_0)
g_2(z_0)})}+\\
&&-\int_{z_0}^z\f{2t^2(t-1)^2f''f'''-2t(t-1)\varphi'f''+\f{d}{dt}[t^2(t-1)^2](f'')^2-4\f{d}{dt}[f'g_1g_2]}{2t(t-1)(t(t-1)f''-\varphi)f''}\,dt+\\
&&-\int_{z_0}^z\left[\f{\varphi'}{t(t-1)f''}\right]\,dt=\\
&&-C_{12}+C_{13}+C_{32}-\ln{(f''(z_0)z_0(z_0-1)-\varphi(z_0))}+\ln{(2\sqrt{f'(z_0)g_1(z_0)
g_2(z_0)})}+\\
&&-\int_{z_0}^z\f{-2t(t-1)\varphi'f''+2\varphi\varphi'}{2t(t-1)(t(t-1)f''-\varphi)f''}\,dt-\int_{z_0}^z\left[\f{\varphi'}{t(t-1)f''}\right]\,dt=\\
&&-C_{12}+C_{13}+C_{32}-\ln{(f''(z_0)z_0(z_0-1)-\varphi(z_0))}+\ln{(2\sqrt{f'(z_0)g_1(z_0)
g_2(z_0)})}
\end{eqnarray*}
This proves that the first equation of the system \eqref{Cij} comes from the first equation of the system \eqref{finalODE3}.
 The remaining equations can be obtained in the same way. 
\newline
\newline
Finally, performing  the substitution $f=\psi+az=\phi=az+b$ with $a=\f{d_{21}^2}{4}$ and $b=-\f{d_{21}d_{23}}{4}$, it is easy to check 
 that the equation \eqref{painlevevi2} reduces to
\eqref{painlevevi3}. Comparing \eqref{painlevevi3} with \eqref{sigmapainleve2},
 we conclude that the equation for $\phi$ and for $\sigma$ coincide iff

\begin{eqnarray*}
\sum_{k=1}^4 v_k^2&=&(2R^2+d_{13}^2-d_{21}d_{13})\\
\sum_{i<j}^4v_{i}^2v_{j}^2&=&
R^4+D(2d_{13}+d_{21})+
\left[\f{d_{13}d_{21}}{2}+\f{d_{23}d_{21}}{2}+d_{21}^2+d^2_{13}+d_{23}d_{13}\right]R^2+\\
&&-\f{d_{21}^2d_{23}^2}{4}+\f{d_{21}^4}{8}+\f{d_{21}^3d_{13}}{4}+\f{d_{21}^2d_{13}d_{23}}{2}+\f{d_{21}^3d_{23}}{2}+\f{d_{21}^2d_{13}^2}{2}\\
\sum_{i<j<k}^4v_i^2v_j^2v_k^2&=&D^2+DR^2(d_{13}+d_{23})-\f{R^4}{4}\left[d_{21}^2+(d_{13}+d_{23})^2\right]
+\f{D}{2}d_{21}^2(d_{13}+d_{23})+\\
&&+\f{R^2}{8}d_{21}^2\left[d_{21}^2
+4d_{13}d_{23}+2d_{13}^2
+2d_{23}^2\right]+\f{d_{21}^4}{16}\left[d_{23}d_{21}
+d_{13}^2+2d_{13}d_{23}\right]\\
(v_1v_2v_3v_4)^2&=&\left[\f{D}{2}d_{21}
+\f{R^2}{4}(d_{13}d_{21}+d_{23}d_{21})+\f{d_{21}^4}{16}+\f{d_{21}^3d_{13}}{8}\right]^2
\end{eqnarray*}
In other words, $v_i^2$ are the roots of the polynomial \eqref{polynomial}.

\endproof

\section{The generalized $\epsilon$-system}

The rotation coefficients
\beq\label{rotmeps}
\beta_{ij}=\f{\prod_{l\ne j}(u^j-u^l)^{\epsilon_l}}{\prod_{l\ne i}(u^i-u^l)^{\epsilon_l}}\f{\epsilon_j}{u^i-u^j}
\eeq
and the Lam\'e coefficients
\beq\label{lamemeps}
H_i=\f{1}{\prod_{l\ne i}(u^i-u^l)^{\epsilon_l}}
\eeq
are solutions of the system (\ref{ED1},\ref{ED2},\ref{ED3},\ref{L1},\ref{L2},\ref{L3}) with
 $d_i=-\sum_{l\ne i}\epsilon^l$.

Thus the associated natural connection $\nabla^{(1)}$ 
\begin{eqnarray*}
\Gamma^{(1)i}_{jk}&=&0\qquad\forall i\ne j\ne k \ne i\\
\Gamma^{(1)i}_{jj}&=&-\Gamma^{(1)i}_{ij}\qquad i\ne j\\
\Gamma^{(1)i}_{ij}&=&\f{\epsilon_j}{u^i-u^j}\qquad i\ne j\\
\Gamma^{(1)i}_{ii}&=&-\sum_{l\ne i}\Gamma^{(1)i}_{li},
\end{eqnarray*}
the associated dual connection $\nabla^{(2)}$ 
\begin{eqnarray*}
\Gamma^{(2)i}_{jk}&=&0\qquad\forall i\ne j\ne k \ne i\\
\Gamma^{(2)i}_{jj}&=&-\f{u^i}{u^j}\Gamma^{(2)i}_{ij}\qquad i\ne j\\
\Gamma^{(2)i}_{ij}&=&\f{\epsilon_j}{u^i-u^j}\qquad i\ne j\\
\Gamma^{(2)i}_{ii}&=&-\sum_{l\ne i}\f{u^l}{u^i}\Gamma^{(2)i}_{li}-\f{1}{u^i},
\end{eqnarray*}
the products $c^i_{jk}=\delta^i_j\delta^i_k$ and $c^{*i}_{jk}=\frac{1}{u^i}\delta^i_j\delta^i_k$,
 the vector fields $e=\sum_{k=1}^n \partial_k$ and $E=\sum_{k=1}^n u^k\partial_k$ define
 a bi-flat semisimple $F$-manifold structure for any choice of $\epsilon_1,\dots,\epsilon_n$.

\subsection{Flat coordinates of the natural connection}

We have to find a basis of flat exact 1-forms $\theta=\theta_i du^i$,
that is, $n$ independent solutions of the linear system of PDEs
\beq
\label{flatform}
\begin{aligned}
&\d_j \theta_i-\f{\epsilon_j\theta_i-\epsilon_i\theta_j}{u^i-u^j}=
0,\,\qquad\,\,i=1,\dots,n,\, j\ne i\\
&\d_i \theta_i+\sum_{k\ne i}\f{\epsilon_i\theta_k-\epsilon_k\theta_i}{u^k-u^i}=0,\,\qquad\,i=1,\dots,n,
\end{aligned}
\eeq  
which is equivalent to
\beq
\label{flatform2}
\begin{aligned}
&\d_j \theta_i-\f{\epsilon_j\theta_i-\epsilon_i\theta_j}{u^i-u^j}=
0,\,\qquad\,\,i=1,\dots,n,\, j\ne i\\
&\sum_{k=1}^n\d_k \theta_i=0,\,\qquad\,i=1,\dots,n.
\end{aligned}
\eeq  
In particular, we have that
$$
0=\sum_{k=1}^n \d_k \theta_i=\sum_{k=1}^n \d_i \theta_k=\d_i\left(\sum_{k=1}^n \theta_k\right),
$$
showing that $\sum_{k=1}^n \theta_k$ is constant if $\theta=\theta_k du^k$ is flat.

A trivial solution of the system (\ref{flatform2}) is given by $\theta_j=\epsilon_j$ for all $j$, corresponding
 to the flat 1-form $\theta^{(1)}=\sum_{l=1}^n \epsilon_ldu^l=df^1$, where
$f^1=\sum_{l=1}^n \epsilon_lu^l$. The other flat coordinates can be chosen according to

\begin{proposition}
 \label{prop:flatcoos}
If $\sum_{l}\epsilon_l\ne 1$, there exist flat coordinates $(f^1,f^2,\dots,f^n)$ such that
$f^p_\epsilon(u)$ is a homogeneous function of degree $(1-\sum_l\epsilon_l)$ for all $p=2,\dots,n$.
 Moreover $e(f^p)=0$ for all $p=2,\dots,n$.
\end{proposition}

The proof works exactly as in the case $\epsilon_i=\epsilon_j$ (see \cite{LP}).
\newline
\newline
For instance, in the case $n=3$ following the same procedure explained in \cite{LP,AL-Reciprocal} one can easily check that
 \beq
 \begin{split}
&f^1=\epsilon_1 u^1+\epsilon_2 u^2+\epsilon_3 u^3\\ 
&f^2=\,{\rm hypergeom}\left([-\f{1}{2}+\f{1}{2}\,\epsilon_1+\f{1}{2}\epsilon_2-\epsilon_3-\f{1}{2}\sqrt {1-\epsilon_1-\epsilon_2}
 \sqrt {-\epsilon_1+8\epsilon_3-\epsilon_2+1},\right.\\
 &\left.-\f{1}{2}+\f{1}{2}\epsilon_1+\f{1}{2}\epsilon_2-\epsilon_3+\f{1}{2}\sqrt {1-\epsilon_1-\epsilon_2}
 \sqrt {-\epsilon_1+8\epsilon_3-\epsilon_2+1}],[-3\epsilon_3+\epsilon_1],1+z \right)+\\ 
 &f^3=
  \left( 1+z \right) ^{1+3\,\epsilon_3-\epsilon_1}{\rm hypergeom}\left( [\f{1}{2}-\f{1}{2}\epsilon_1+\f{1}{2}\epsilon_2+2\epsilon_3-\f{1}{2}\sqrt {1-\epsilon_1
 -\epsilon_2}\sqrt {-\epsilon_1+8\epsilon_3-\epsilon_2+1},\right.\\
 &\left.+\f{1}{2}-\f{1}{2}\epsilon_1+\f{1}{2}
 \epsilon_2+2\epsilon_3+\f{1}{2}\sqrt {1-\epsilon_1-\epsilon_2}
 \sqrt {-\epsilon_1+8\epsilon_3-\epsilon_2+1}],[2+3\epsilon_3-\epsilon_1],1+z \right)
 \end{split}
 \eeq
 where $z=\f{u^3-u^2}{u^2-u^1}$.

\subsection{Principal hierarchy}
Given an $F$-manifold with compatible \emph{flat} connection one can construct  a hierarchy of integrable
 quasilinear PDEs called \emph{principal hierarchy} \cite{LPR}. It is defined in the
  following way, which is a straightforward generalization of the original definition
   given by Dubrovin in the case of Frobenius manifolds \cite{du93}.

First of all, one defines the so-called \emph{primary flows}:
\begin{equation}\label{primflo}
u^i_{t_{(p,0)}}=c^i_{jk}X^k_{(p,0)}u^j_x,
\end{equation}  
where  $(X_{(1,0)},\dots, X_{(n,0)})$ is a basis of flat vector fields. Then,  
starting from these flows, one can define the ``higher flows'' of the hierarchy,
\beq
\label{hiflows}
u^i_{t_{(p,\alpha)}}=c^i_{jk}X^k_{(p,\alpha)}u^j_x,
\eeq
by means of the following recursive relations:
\beq\label{recrel}
\nabla_j X^i_{(p,\alpha)}=c^i_{jk}X^k_{(p,\alpha-1)}.
\eeq
In this section we will study the principal hierarchy
  associated with the bi-flat $F$-manifold defined above. One of the flows is the generalized $\epsilon$-system \cite{pavlov}.
\newline
\newline
\n
{\bf The primary flows}. In order to define the primary flows we need a frame of flat vector fields
$X=X^i\frac{\partial}{\partial u^i}$,
 that is, $n$ independent solutions of the linear system of PDEs
\beq
\begin{aligned}
\label{flatvf}
&\d_j X^i+\f{\epsilon_jX^i-\epsilon_iX^j}{u^i-u^j}=
0,\,\qquad \,i=1,\dots,n,\,j\ne i\\
&\d_i X^i-\sum_{k\ne i}\f{\epsilon_iX^k-\epsilon_kX^i}{u^k-u^i}=0,\,\qquad \,i=1,\dots,n
\end{aligned}
\eeq
which is equivalent to
\begin{eqnarray}
&&\label{E1}\d_j X^i+\f{\epsilon_jX^i-\epsilon_iX^j}{u^i-u^j}=
0,\,\qquad \,i=1,\dots,n,\, j\ne i\\
&&\label{E2}[e,X]=0.
\end{eqnarray}
Comparing (\ref{flatvf}) with (\ref{flatform}), one notices that the components $X^i$ of a flat vector fields for $(\epsilon_1,\dots,\epsilon_n)$ are given by the components of a flat 1-form for $(-\epsilon_1,\dots,-\epsilon_n)$.
\newline
\newline
\n
{\bf The higher flows}. In the case of generalized $\epsilon$-system, the system \eqref{recrel} is equivalent to the system 
\begin{eqnarray}
\label{HF1}&&\d_j X^i_{(p,\alpha)}+\epsilon\f{X^i_{(p,\alpha)}-X^j_{(p,\alpha)}}{u^i-u^j}=
0,\,\qquad\,i=1,\dots,n,\,j\ne i\\
\label{HF2}&&[e,X_{(p,\alpha)}]=X_{(p,\alpha-1)}.
\end{eqnarray}
Since locally  $X^i_{(p,\alpha)}=
\d_i K_{(p,\alpha)}$ (the functions $K_{(p,\alpha)}$ are the coefficients of the deformed flat coordinates for the generalized $\epsilon$-system with
 $\epsilon_i\to-\epsilon_i$) the system \eqref{HF1}
 can be written as
\beq
\label{flatcoo1}
(u^i-u^j)\d_j \d_i K_{(p,\alpha)}+(\epsilon_j\d_i K_{(p,\alpha)}-\epsilon_i\d_j K_{(p,\alpha)})=
0,\,\qquad\,\,i=1,\dots,n,\, j\ne i
\eeq
or in compact form as
\begin{equation*}
dd_L K_{(p,\alpha)} =dK_{(p,\alpha)}\wedge df^1,
\end{equation*}
where $f^1=\sum_l\epsilon_lu^l$ and $d_L$ is the differential associated with the torsionless
 tensor field $L^i_j=u^i\delta^i_j$ \cite{FN}.  
 This is a crucial remark because \eqref{flatcoo1}
 can be recursively solved by
$$dK_{(p,\alpha)}=d_L K_{(p,\alpha)}-K_{(p,\alpha)}df^1.$$
 Using this fact, 
it is easy to check that ---apart from some critical values of $\epsilon_i$--- the functions $K_{(p,\alpha)}$
 obtained in this  way (properly normalized) provide the solutions
   of the full system
 (\ref{HF1},\ref{HF2}). 
 
\begin{proposition}
Suppose that $\sum_l\epsilon_l\ne -1$
 and let $\left(f^1=\sum_l\epsilon_lu^l,f^2,\dots,f^n\right)$ be the flat coordinates
 of the natural connection of the $(-\epsilon_1,\dots,-\epsilon_n)$-system described in Proposition
 \ref{prop:flatcoos}.
If $K_{(p,\alpha)}$ are the functions defined recursively by
\beq\label{recak}
K_{(p,0)}=f^p,\,\,\,dK_{(p,\alpha+1)}=d_L K_{(p,\alpha)}-K_{(p,\alpha)} df^1,
\qquad\,\,\,\alpha\ge 0,
\eeq
and
\begin{equation}\label{recrel2}
Y_{(p,\alpha)}^i=-\frac1{\epsilon_i}\d_i K_{(p,\alpha)},
\qquad \,\,\,\alpha\ge 0,
\end{equation}
then the vector fields $X_{(1,\alpha)}=\frac1{\prod_{j=1}^\alpha(j-\sum_l\epsilon_l)}Y_{(1,\alpha)}$ (for $\sum\epsilon_l\ne j$ with
$j=1,\dots,\alpha$) and  $X_{(p,\alpha)}=\frac1{\alpha!}Y_{(p,\alpha)}$, for $p=2,\dots,n$, satisfy the recursion relations
(\ref{recrel}).
\newline
Moreover the recursion relations (\ref{recak}) are algebraically solved by
\beq\label{fid}
K_{(1,\alpha)}=\f{1}{\alpha+1}
\left[\sum_{l=1}^n (u^l)^2\d_l K_{(1,\alpha-1)}-\left(\sum_{l=1}^n\epsilon_l u^l\right) K_{(1,\alpha-1)}\right]
\eeq  
and, for $\alpha\ne -1-\sum_l\epsilon_l$, by
\beq\label{fid2}
K_{(p,\alpha)}=\f{1}{\alpha+1+\sum_l \epsilon_l}
\left[\sum_{l=1}^n (u^l)^2\d_l K_{(p,\alpha-1)}-\left(\sum_{l=1}\epsilon_l u^l\right) K_{(p,\alpha-1)}\right]
,\qquad p=2,\dots,n.
\eeq  
\end{proposition}

\n
The proof works as in the case $\epsilon_i=\epsilon_j$  which is treated with details in \cite{LP}.  

\begin{remark}
The vector fields $Y_{(p,\alpha)}$ \eqref{recrel2} define the twisted Lenard-Magri chain 
 \cite{AL-LMchains} associated to the almost hydrodynimically connections $\nabla^{(1)}$
 and $\nabla^{(3)}$: 
$$\Gamma^{(3)i}_{jk}=\Gamma^{(2)i}_{jk}+(1-\sum_l\epsilon_l)c^{*i}_{jk}=
\Gamma^{(2)i}_{jk}+(1-\sum_l\epsilon_l)\f{1}{u^i}\delta^i_j\delta^i_k.$$
This means that they satisfy the following recursive relations
$$d_{\nabla^{(1)}}Y_{(n,\alpha)}=d_{\nabla^{(3)}}\left(E\circ Y_{(n-1,\alpha)}\right),$$
as one can easily verify by straightforward computation. This means that the recursive 
 procedure to construct integrable hierarchies based on the Fr\"olicher-Nijnhuis theory 
 \cite{LM,L2006} is a particular case of the more general setting developed in \cite{AL-LMchains}.
\end{remark}

\begin{remark}
For generic values of $\epsilon_1,\dots,\epsilon_n$
 the principal hierarchy is not hamiltonian w.r.t. a local
 Poisson bracket of hydrodynamic type. However according to \cite{AL-Poisson} any 
 flow can be written as
$$u^i_t=P^{ij}\alpha_j$$ 
where $\alpha$ is a \emph{non exact} $1$ form,
$$P^{ij}=g^{ij}\d_x-g^{il}\Gamma^j_{lk}u^k_x$$
is the local Poisson bivector of hydrodynamic type 
 associated to a flat metric $g$  compatible with the natural connection: $\nabla^{(1)} g=0$.
\end{remark}
\subsection{Reciprocal transformations}
To conclude this Section we apply the results of \cite{AL-Reciprocal} to the generalized $\epsilon$-system.
\begin{theorem}
Suppose $\beta_{ij}$ satisfies system (\ref{ED1},\ref{ED2},\ref{ED3}) and $H_i$ satisfies the corresponding system (\ref{L1},\ref{L2}).
 Assume that $A$ is a homogeneous flat coordinate of degree $k$ of the natural connection satisfying the condition $e(A)=0$, then 
\beq\label{newRC}
\tilde \beta_{ij}:=\beta_{ij}-\f{H_i}{H_j}\d_j \ln(A), \quad i\neq j,
\eeq
and
\beq\label{newLC}
\tilde H_{i}:=\f{H_i}{A},
\eeq
satisfy systems (\ref{ED1},\ref{ED2},\ref{ED3}) and (\ref{L1},\ref{L2})  respectively, with $d_i$ replaced by $d_i-k$ in \eqref{L3}. 
\end{theorem}

In the case $d_i=d_j$ the proof was given in \cite{AL-Reciprocal}. The general case is completely similar.

Since $n-1$ flat coordinates of the generalized $\epsilon$-system satisfy the hypothesis 
 of the above theorem with $k=1-\sum_l\epsilon_l$,  we have immediately the following corollary.

\begin{corollary}
Let $\beta_{ij}$  be the rotation coefficients  \eqref{rotmeps} and $H_i$ the Lam\'e coefficients \eqref{lamemeps},
 then the new rotation coefficients \eqref{newRC} and the new Lam\'e coefficents \eqref{newLC} with $A=f^k,\,k=2,\dots,n$ 
 define a new solution
 of systems (\ref{ED1},\ref{ED2},\ref{ED3}) and (\ref{L1},\ref{L2}) with $d_i$ replaced by $d_i-1+\sum_l\epsilon_l$. 
\end{corollary}

In other words, using the language of \cite{AL-Reciprocal},
 the reciprocal $F$-manifold  associated with any flat coordinates $f^2,\dots,f^n$ is still a bi-flat
 $F$-manifold.

\subsection*{Acknowledgments}
I thank Alessandro Arsie for many fruitful discussions.

\end{document}